\documentclass{llncs}
\usepackage{amssymb}
 \usepackage{graphicx}
 \usepackage{fancyhdr} 
  \usepackage{soul, color}
  \sethlcolor{green}
 \usepackage[neveradjust]{paralist}
\usepackage{xspace}
\usepackage{latexsym}

\newcommand{\dlrifd}{\mathcal{DLR}_{\mbox{\emph{{\fontfamily{phv}\selectfont{\scriptsize ifd}}}}}}

\def\I   {{\ensuremath{\mathcal{I}}\xspace}}    

\newtheorem{thm}{\textsc{Theorem}}
\newtheorem{ex}{\textbf{Example}}

\begin{document}

\title{Mapping the Object-Role Modeling language ORM2 into Description Logic language $\mathcal{DLR}_{ifd}$\thanks{This is an updated version of {\em Keet, C.M. Mapping the Object-Role Modeling language ORM2 into Description Logic language {\em $\dlrifd$}. KRDB Research Centre Technical Report KRDB07-2, Free University of Bozen-Bolzano, 15-2-2007. http://arxiv.org/abs/cs.LO/0702089v1}: it discusses additional recent literature, has the ORM figures made with the new tool NORMA, makes the mappable elements more readable (including fixing some typos in the mappings and adding more explanations), and has the proofs of correctness of encoding.}}
\author{C. Maria Keet}
\institute{Faculty of Computer Science, Free University of Bozen-Bolzano, Italy\\ 
\email{keet@inf.unibz.it}}
\maketitle
\begin{abstract}
In recent years, several efforts have been made to enhance conceptual data modelling with automated reasoning to improve the model's quality and derive implicit information. One approach  to achieve this in implementations, is to constrain the language. Advances in Description Logics can help choosing the right language to have greatest expressiveness yet to remain within the decidable fragment of first order logic to realise a workable implementation with good performance using DL reasoners. The best fit DL language appears to be the ExpTime-complete $\dlrifd$. To illustrate trade-offs and highlight features of the modelling languages, we present a precise transformation of the mappable features of the very expressive (undecidable) ORM/ORM2 conceptual data modelling languages to exactly $\dlrifd$. Although not all ORM2 features can be mapped, this is an interesting fragment because it has been shown that $\dlrifd$ can also encode UML Class Diagrams and EER, and therefore can foster interoperation between conceptual data models and research into ontological aspects of the modelling languages. 
\end{abstract}


\section{Introduction}
\label{sec:intro}
Various strategies and technologies are being developed to reason over conceptual data models to meet the same or slightly different requirements and aims. An important first distinction is between the assumption that modellers should be allowed to keep total freedom to model what they deem necessary to represent and subsequently put constraints on which parts can be used for reasoning or accept slow performance versus the assumption that it is better to constrain the language {\em a priori} to a subset of first order logic so as to achieve better performance and a guarantee that the reasoner terminates. The former approach is taken by \cite{Queralt08} using a dependency graph of the constraints in a UML Class Diagram + OCL and by first order logic (FOL) theorem provers. The latter approach is taken by \cite{Smaragdakis07,Kaneiwa06,Cabot08,Cadoli07,Calvanese98,Artale07er,Jarrar07,Franconi00,Keet07dl}  who experiment with different techniques. For instance, \cite{Smaragdakis07,Kaneiwa06} use special purpose reasoners for ORM and UML Class Diagrams, \cite{Cabot08,Cadoli07} encode a subset of UML class diagrams as a Constraint Satisfaction Problem, and \cite{Calvanese98,Artale07er,Jarrar07,Franconi00,Keet07dl} use a Description Logic (DL) framework for UML Class Diagrams, ER, EER, and ORM. Of the many DL languages experimented with, $\dlrifd$ provides the best trade-off between expressiveness and available features for conceptual data modelling languages for UML Class Diagrams without OCL and the less expressive EER. However, one would also want to include at least some OCL or cater for a richer language such as Object-Role Modeling. In fact, whereas for some constraints in the UML framework one has to resort to OCL, with ORM there are still icons in the graphical language. In addition, ORM has explicitly external uniqueness, which is a more general version of UML's qualified association, and role subsetting, which corresponds to subsetting  of UML's association ends, forcing one to add a mapping whereby we thus easily can add this also to the UML to $\dlrifd$ mapping of \cite{Berardi05} so as to be more truthful to the UML specification \cite{UMLSpec07}. In addition, ORM is a so-called ``true'' conceptual modelling language in the sense that it is independent of the application scenario and it has been mapped into both UML class diagrams and ER \cite{Halpin01}. That is, ORM and its successor ORM2 can be used in the conceptual analysis stage for database development, application software development, requirements engineering, business rules, and other areas \cite{Halpin01,Balsters06,Evans05,Halpin08,Pepels05}. Thus, if there is an ORM-to-DL mapping, the possible applications for automated reasoning services can be greatly expanded. Furthermore, given that EER and a restricted version of UML Class Diagrams also have $\dlrifd$-encoding, this would greatly simplify model interoperability and mutual benefit of each language's strengths. Therefore, our aim is to map (most of) ORM and its successor, ORM2, into $\dlrifd$; the mappable constraints and correctness of encoding will be presented in section \ref{app:ormdlrifd}. This will also give a clear view on trade-offs between DL languages and the choice for using DL for automated reasoning over conceptual models. The reader may be aware of previous work by Jarrar \cite{Jarrar07} that attempted to complete the same task.  The two main problems with that work is that several of his mapping ``rules'' did not remain within $\dlrifd$ but various constructors and features were borrowed from other DL languages (hence, the mapping in toto is not to any particular DL language) and a few ORM constraints were incorrect or incomplete. We shall discuss these issues, as well as some general reflections, in section \ref{sec:disc}. Finally, we close with conclusions in section \ref{sec:concl}.

\section{The $\mathcal{DLR}_{ifd}$ language}
\label{sec:dlrifd}
Description Logics (DL) languages are decidable fragments of first order logic and are used for logic-based knowledge representation, such as conceptual modelling and ontology development. The basic ingredients of all DL languages are \textit{concepts} and \textit{roles}, where a DL-role is an $n$-ary predicate ($n \geq 2$) and several constructs, thereby giving greater or lesser expressivity and efficiency of automated reasoning over a logical theory. DL knowledge bases are composed of the \emph{Terminological Box} (TBox), which contains axioms at the concept-level, and the \emph{Assertional Box} (ABox) that contains assertions about instances. A TBox corresponds to a formal conceptual data model or can be used to represent a type-level ontology; refer to \cite{Baader03} for more information about DLs and their usages. For formal conceptual data modelling, we use $\dlrifd$ \cite{Calvanese01}. This DL language was developed to provide a formal characterization of conceptual modelling languages to enable automated reasoning over the conceptual models to improve their quality and that of the resulting software application, and to use it as unifying paradigm for database integration through integrating their respective conceptual models \cite{Calvanese98a}. Take atomic relations (\textbf{P}) and atomic concepts \emph{A} as the basic elements of $\dlrifd$, which allows us to construct arbitrary relations (arity $\geq 2$) and arbitrary concepts according to the following syntax:\\
\indent \textbf{R} $\longrightarrow \top_n | $ \textbf{P} $ | \mbox{ } (\$ i / n: C) \mbox{ } |\mbox{ } \neg$\textbf{R} $|$ \textbf{R}$_1 \sqcap$ \textbf{R}$_2$ \\
 \indent $C \longrightarrow \top_1 |$ $A$ $|$ $\neg C$ $|$ $C_1 \sqcap C_2$ $|$ $\exists[\$ i]$\textbf{R} $|$ $\leq k [\$ i]$\textbf{R}\\
\noindent where $i$ denotes a component of a relation (the equivalent of an ORM-role); if components are not named, then integer numbers between 1 and $n_{max}$ are used, where $n$ is the arity of the relation; $k$ is a nonnegative integer for cardinality constraints. Only relations of the same arity can be combined to form expressions of type \textbf{R}$_1 \sqcap$ \textbf{R}$_2$, and $i \leq n$, \emph{i.e.}, the concepts and relations must be well-typed. Further, $\dlrifd$ has \emph{i}dentification assertions on a concept $C$, which has the form (\textbf{id} $C [i_1]R_1, ..., [i_h]R_h$), where each $R_j$ is a relation and each $i_j$ denotes one component of $R_j$. Then, if $a$ is an instance of $C$ that is the $i_j$-th component of a tuple $t_j$ of $R_j$, for $j \in \{1,...,h \}$, and $b$ is an instance of $C$ that is the $i_j$-th component of a tuple $s_j$ of $R_j$, for $j \in \{1, ..., h\}$, and for each $j$, $t_j$ agrees with $s_j$ in all components different from $i_j$, then $a$ and $b$ are the same object \cite{Calvanese01}. This gives greater flexibility how to identify DL-concepts, most notably external uniqueness (/weak entity types/qualified associations \cite{Halpin01}), and objectification. Second, $\dlrifd$ has non-unary \emph{f}unctional \emph{d}ependency assertions on a relation $R$, which has the form (\textbf{fd} $R$ $i_1, ..., i_h \rightarrow j$), where $h \geq 2$, and $i_1, ..., i_h, j$ denote components of $R$ (unary \textbf{fd}s lead to undecidability \cite{Calvanese01}) and for all $t, s \in R^{\cal{I}}$, we have that $t[r_1] = s[r_1], ... , t[r_i] = s[r_i] \mbox{ implies } t_j = s_j$. This is useful primarily for UML class diagram's methods and ORM's derived-and-stored fact types. 
The model-theoretic semantics of $\cal{DLR}$ is specified through the usual notion of interpretation, where $\cal{I}$$= (\Delta ^{\cal{I}}, \cdot ^{\cal{I}})$, and the interpretation function $\cdot ^{\cal{I}}$ assigns to each concept $C$ a subset $C^{\cal{I}}$ of $\Delta^{\cal{I}}$ and to each $n$-ary \textbf{R} a subset \textbf{R}$^{\cal{I}}$ of $(\Delta ^{\cal{I}})^n$, such that the  conditions are satisfied following Table \ref{tab:sem}. 
Observe that {\bf id} and {\bf fd} are not mentioned in the semantics for $\dlrifd$ in Table \ref{tab:sem}: there are no changes in semantic rules because the algorithms for them are checked against a (generalized) ABox \cite{Calvanese01}. 
\begin{table}[h]
	\centering
	\caption{Semantics of $\dlrifd$.}
\begin{tabular}{cc}
\hline
$\top ^{\cal{I}}_n  \subseteq  (\Delta ^{\cal{I}})^n$  &  $A^{\cal{I}}  \subseteq  \Delta ^{\cal{I}}$ \\
$\mbox{\textbf{P}}^{\cal{I}}  \subseteq  \top ^{\cal{I}}_n$ & $ (\neg C)^{\cal{I}}   =  \Delta ^{\cal{I}} \setminus C^{\cal{I}}$ \\
$(\neg \mbox{\textbf{R}})^{\cal{I}}  =  \top ^{\cal{I}}_n \setminus \mbox{\textbf{R}}^{\cal{I}}$ & $(C_1 \sqcap C_2)^{\cal{I}}  =  C_1^{\cal{I}} \cap C_2^{\cal{I}}$ \\ 
$(\mbox{\textbf{R}}_1 \sqcap \mbox{\textbf{R}}_2 )^{\cal{I}}  =  \mbox{\textbf{R}}_1^{\cal{I}} \cap \mbox{\textbf{R}}_2^{\cal{I}}$ & $(\$ i / n : C)^{\cal{I}}  =  \{(d_1, ..., d_n) \in \top ^{\cal{I}}_n | d_i \in C^{\cal{I}} \}$ \\ 
 $  \top ^{\cal{I}}_1   =  \Delta ^{\cal{I}}$ & $  (\exists [\$ i]\mbox{\textbf{R}})^{\cal{I}}  =  \{ d \in \Delta ^{\cal{I}} | \exists(d_1,...,d_n) \in \mbox{\textbf{R}}^{\cal{I}}.d_i =d \} $  \\ 
 & $ (\leq k [\$ i]\mbox{\textbf{R}})^{\cal{I}}  =  \{ d \in \Delta ^{\cal{I}} || \{ (d_1,...,d_n) \in \mbox{\textbf{R}}^{\cal{I}}_1 | d_i =d | \} \leq k \}$ \\ \hline
\end{tabular}
\label{tab:sem}
\end{table}
\vspace{-3mm}

A {\em knowledge base} is a finite set $\cal{KB}$ of $\dlrifd$ axioms of the form $C_{1}\sqsubseteq C_{2}$ and $R_{1} \sqsubseteq R_{2}$. An interpretation $\I$ satisfies $C_1 \sqsubseteq C_2$ ($R_{1} \sqsubseteq R_{2}$) if and only if the interpretation of $C_1$ ($R_1$) is included in the interpretation of $C_2$ ($R_2$), i.e. $C_1^{\mathcal{I}} \subseteq C_2^{\mathcal{I}}$ ($R_1^{\mathcal{I}} \subseteq R_2^{\mathcal{I}}$). $\top _1$ denotes the interpretation domain, $\top _n$ for $n \geq 1$ denotes a subset of the $n$-cartesian product of the domain, which covers all introduced $n$-ary relations; hence ``$\neg$'' on relations means difference rather than the complement. The $(\$ i / n : C)$ denotes all tuples in $\top _n$ that have an instance of $C$ as their $i$-th component. The following abbreviations can be used: $\bot$ for $\neg \top_1$, $C_1 \sqcup C_2$ for $\neg (\neg C_1 \sqcap \neg C_2)$, $C_1 \Rightarrow C_2$ for $\neg C_1 \sqcup C_2$, $(\geq k [i]R)$ for $\neg (\leq k-1 [i]R)$, $\exists [i]R$ for $(\geq 1 [i]R)$, $\forall [i]R$ for $\neg \exists [i]\neg R$, $R_1 \sqcup R_2$ for $\neg (\neg R_1 \sqcap \neg R_2)$, and $(i : C)$ for $(i / n : C)$ when $n$ is clear from the context. Note that for a qualified role, as in $\exists P.C$ and represented in $\dlrifd$ as $\exists [\$1](P \sqcap (\$ 2 / 2 : C))$, its inverse, $\exists P^-.C$, is represented as $\exists [\$2](P \sqcap (\$ 1 / 2 : C))$, likewise for universal quantification ($\forall P.C$ as $\neg \exists [\$1](P \sqcap (\$ 2 / 2 : \neg C))$ and its inverse $\forall P^-.C$ as $\neg \exists [\$2](P \sqcap (\$ 1 / 2 : \neg C))$ (\cite{Baader03} chapter 5).

\section{ORM2 to {\em $\dlrifd$} transformation} 
 \label{app:ormdlrifd}
 
After a brief ORM introduction and the technical preliminaries (section \ref{sec:prelim}), the mappable elements of ORM2 into $\dlrifd$ are listed in section \ref{sec:mappable}, which is followed by the proof of correctness of encoding.

\subsection{Brief informal overview of ORM}

\begin{figure}[h]
	\centering
		\includegraphics[width=0.90\textwidth]{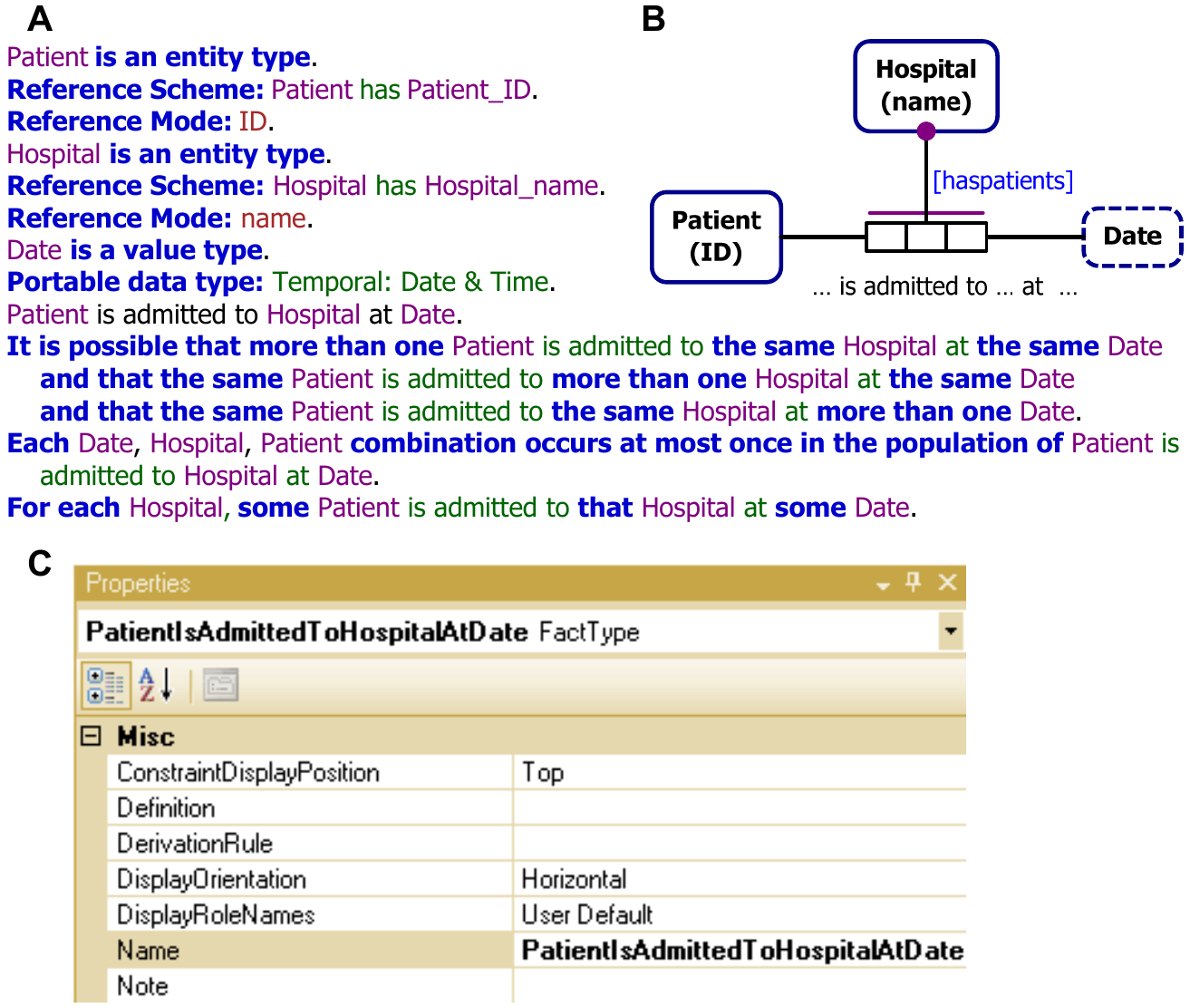}
	\caption{A: verbalization of the small ORM2 conceptual model, consisting of one fact type; B: graphical depiction of the ORM2 conceptual model, depicting two object types, a value type, a ternary relation, label for the reading, name of the first role in ``[ ]'', mandatory (blob) and uniqueness (line) constraints; C: properties box of the fact type, displaying the name of the relation}
	\label{fig:PatientEx}
\end{figure}

\begin{figure}[h]
	\centering
		\includegraphics[width=1.00\textwidth]{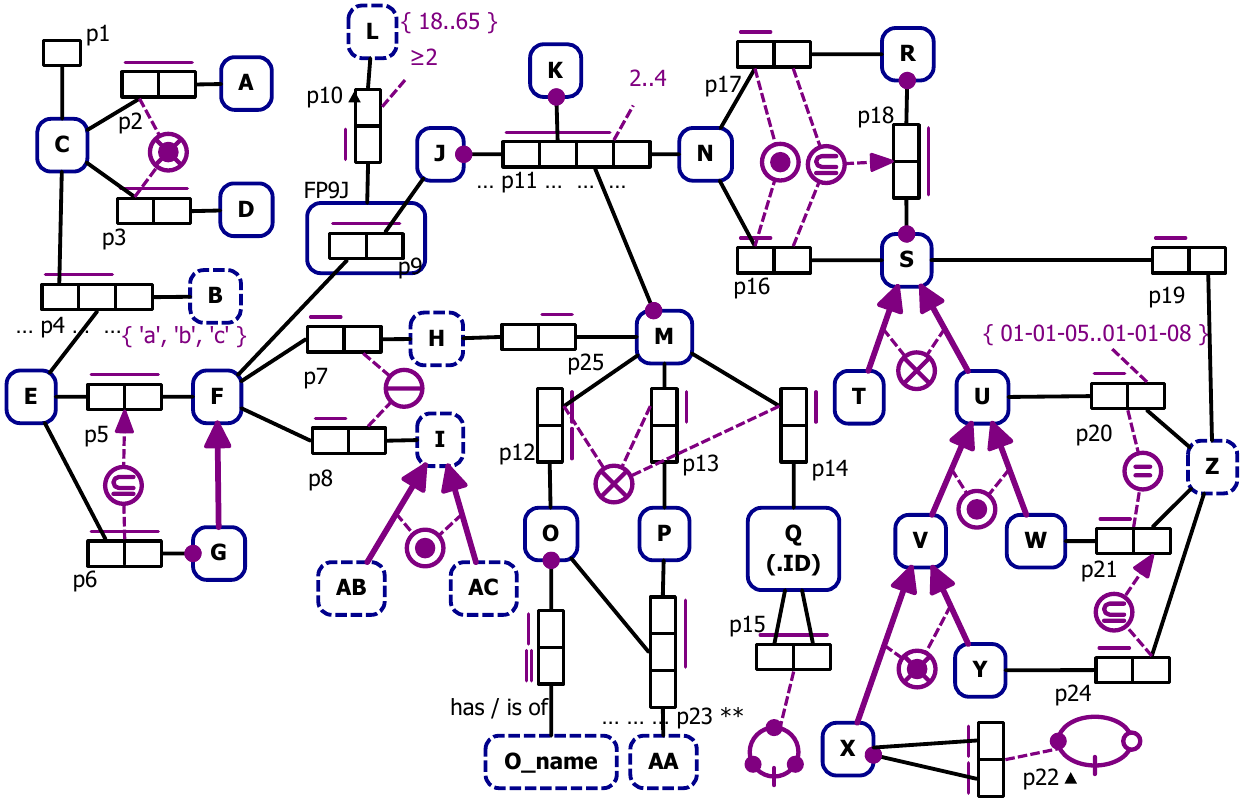}
	\caption{ORM2 model with most constraints in most of the allowable combinations. All object types should have a reference scheme like $O$ and $Q$, where $Q$ has the default notation to unclutter the diagram and $O$ shows the expanded full representation.}
	\label{fig:ormconstraints}
\end{figure}

The basic building blocks of the ORM language are \emph{object type} (class), \emph{value type} (data type), 
\emph{fact type} (typed relation), \emph{role}---that what the object or value type `plays' in the relation---and a plethora of \emph{constraints}. ORM supports $n$-ary relations, where $n \geq 1$, 
and this $n$-ary predicate, $R$, is composed of $r_1, ..., r_n$ roles where each role has one object type, denoted with $C_1, ..., C_n$, associated with it. Roles and predicates are globally unique, even though the `surface labeling' by the modeller or domain expert may suggest otherwise. An example of a fact type is shown in Fig. \ref{fig:PatientEx}, which was made with the NORMA CASE tool [http://sourceforge.net/projects/orm/]. 
The diagrammatic representation has
\begin{compactenum}[1)]
\item the name of the relation, which is displayed in the properties box of the relation and is, in this example, 
generated automatically by the software and called {\fontfamily{phv}\selectfont{\small PatientIsAdmittedToHospitalAtDateDate}}; 
\item role names, such as the manually named {\fontfamily{phv}\selectfont{\small [haspatients]}} for the the role that object type {\fontfamily{phv}\selectfont{\small Hospital}} plays; and 
\item a label attached to the relation, ``{\fontfamily{phv}\selectfont{\small ... is admitted to ... at date ...}}'', which is used for the verbalization by filling the ellipses with the names of the participating object types ({\fontfamily{phv}\selectfont{\small Patient is admitted to Hospital at date Date}});
\item two object types, {\fontfamily{phv}\selectfont{\small Patient}} and {\fontfamily{phv}\selectfont{\small Hospital}}, and a value type {\fontfamily{phv}\selectfont{\small Date}};
\item a \emph{reference scheme} for each object type, shown in compact format with {\fontfamily{phv}\selectfont{\small (ID)}} for {\fontfamily{phv}\selectfont{\small Patient}} and {\fontfamily{phv}\selectfont{\small (name)}} for {\fontfamily{phv}\selectfont{\small Hospital}};
\item spanning uniqueness constraint drawn with a line next to the roles included in the uniqueness constraint, in this case all three roles are included;
\item mandatory participation of {\fontfamily{phv}\selectfont{\small Hospital}} in the fact type, denoted with a blob.
\end{compactenum}
Many more features are available in the language than are illustrated in this example. We deal with them in the next subsection and Fig. \ref{fig:ormconstraints}.

ORM diagrams can be transformed more or less into, among others, ER and UML Class Diagrams, IDEFX1, SQL table definitions, C, Visual Basic, and XML. More information on these modelling, design- and implementation-oriented transformations can be found in, e.g., \cite{Halpin01,Halpin08}. 

\subsection{Preliminaries}
\label{sec:prelim}   
The here presented transformation assesses all components and constraints of ORM2, hence also of ORM, except deontic constraints that were recently added to ORM2 (compared to ORM in \cite{Halpin89}, ORM2 also supports exclusive total covering of subtypes, role values, and deontic constraints). As starting point, we used the ORM formalisation by Halpin \cite{Halpin89}, where available, which was the first formalisation of ORM. 
Other formalizations of ORM \cite{Hofstede93,Hofstede98} do not differ significantly from Halpin's version. They make clearer distinctions between roles and predicates and the relation between them and the naming versus labeling of ORM elements, but they cover fewer constraints.  
In the following, we take this same unambiguous approach of \cite{Hofstede93,Hofstede98}. That is, any ORM model has for each predicate a \emph{surface reading label}, such as ``{\fontfamily{phv}\selectfont{\small ...admitted to...at date...}}'' in Fig. \ref{fig:PatientEx}, a \emph{predicate name}, which could be $hospitalAdmission$ that is formally typed as $\forall x,y,z (hospitalAdmission(x,y,z) \rightarrow Patient(x) \land Hospital(y) \land Date(z))$, and each of the roles can bear a name, or be simply indexed from left to right or top to bottom ($r_1$, $r_2$, and $r_3$) throughout the model (all roles and predicates are uniquely identified). An important consequence of such a commitment concerns the customary ORM practice of providing ``forward'' and ``backward'' reading labels so as to make nice pseudo-natural language sentences that can be verified by the domain expert; for instance, with a label {\fontfamily{phv}\selectfont{\small orders / orderedBy}}, one then reads a fact type as {\fontfamily{phv}\selectfont{\small Customer orders Book}} and {\fontfamily{phv}\selectfont{\small Book orderedBy Customer}}. These, however, are reading labels and do not necessarily imply that $orders(x,y)$ has $orderedBy(y,x)$ as its inverse relation; in fact, that particular predicate may well be named $ordering$ and the reading labels are just that. Ter Hofstede and Proper \cite{Hofstede98} refer to the distinction between predicate and role names versus reading labels as \emph{deep semantics} versus \emph{surface semantics}; for the ORM2 to $\dlrifd$ transformation, we are interested in the former.

\subsection{Mappable elements and constraints}
\label{sec:mappable}

The text in boldface in the following list indicates the name of the element or constraint, followed by the FOL characterisation taken from or based on \cite{Halpin89}, a ``mapped to'', with the $\dlrifd$ representation and a reference to Fig. \ref{fig:ormconstraints} for its graphical notation in the NORMA case tool. At times, we will use abbreviations in the FOL:  to shorten representation of an $n$-ary relation, we use an underlined variable, as in \underline{\emph{x}}, which is an abbreviation for a sequence $x_1, ... , x_n$ in an $n$-ary relation, and sometimes we use this when we need access to one or two of the participating variables in the predicate, then the sequence comprises  $x_1, ... , x_{n-2}$ (the difference is clear from the context).
\begin{compactenum}[1.]
\item \textbf{Object type}, $\forall x C(x)$\\
	 {\em mapped to}: $C$\\
	 {\em example}:  the solid roundtangles such as {\fontfamily{phv}\selectfont{\small A}} 
\item \textbf{Named value type} (data type or lexical type), which permits values of some set $\{v_1, ..., v_n\}$ where the values of $C$ are not constrained to specific values, and the value type $C$, thus $\forall x (C(x) \equiv x \in \{v_1, ..., v_n\})$\\
	{\em mapped to}: $C$, the concrete domain ($T$) of the value type can be a user-defined or built-in one, such as \texttt{String} and \texttt{Integer}; we then have, like in \cite{Berardi05} p97, that a relation, $R$, from $C$ to concrete domain of type $T$ is represented as $C \sqsubseteq \forall [r1](R \Rightarrow (r2:T))$ \\
	{\em example}: the dashed roundtangles such as {\fontfamily{phv}\selectfont{\small B}}, {\fontfamily{phv}\selectfont{\small H}}, and {\fontfamily{phv}\selectfont{\small I}} (the domain is shown only in the properties box). 
\item \textbf{Unary relation}, $\forall x (R(x) \rightarrow C_i(x))$\\ 
	 {\em mapped to}: $R \sqsubseteq (r_i : C_i) \sqcap (r_j : C^{\prime})$ where $C^{\prime}$ is a auxiliary new introduced filler object- or value type for the other position in the relation\\
	 {\em example}: role {\fontfamily{phv}\selectfont{\small p1}} that is connected to {\fontfamily{phv}\selectfont{\small C}}.
\item \textbf{Binary relation},  $\forall x, y (R(x,y) \rightarrow C_i(x) \land C_j(y))$\\
	 {\em mapped to}: $R \sqsubseteq (r_i : C_i) \sqcap (r_j : C_j)$ \\
	 {\em example}: {\fontfamily{phv}\selectfont{\small p2}} relating object types {\fontfamily{phv}\selectfont{\small C}} and {\fontfamily{phv}\selectfont{\small A}}.
\item \textbf{$n$-ary relation}, $\forall x_1,..., x_n (R(x_1, ..., x_n) \rightarrow C_1(x_1) \land ... \land C_n(x_n))$ where that $C_1,...,C_n$ may be object types or value types\\
	 {\em mapped to}: $R \sqsubseteq (r_1 : C_1) \sqcap ... \sqcap (r_n : C_n)$ or, in short, $R \sqsubseteq \sqcap_{i=1}^n (r_i : C_i)$, hence a generalisation of the previous two\\
	 {\em example}: any relation {\fontfamily{phv}\selectfont{\small p1}}, ..., {\fontfamily{phv}\selectfont{\small p24}} and the `hidden' relations for the reference schemes (expanded for object type {\fontfamily{phv}\selectfont{\small O}}).
\item \textbf{Object type} participating in an $n$-ary relation $\forall x$\underline{\emph{z}} $C(x) \rightarrow R(x,$\underline{\emph{z}}$)$\\
 {\em mapped to}: $C \sqsubseteq \forall [r_i]R$\\
 {\em example}: all object types in the figure participate in at least one relation.
\item \textbf{Named value type}, where the values of the concrete domain of the value type are \emph{constrained} to specific values $\{v_1, ..., v_i\}$, and value type $C$ with $\forall x (C(x) \equiv x \in \{v_1, ..., v_i\})$\\ 
	{\em mapped to}: $C_i \sqsubseteq$ $\forall [r_i]R(R \Rightarrow (r_j : T_j) \sqcap (T_j \equiv \{v_1, ..., v_i\})$  s.t. for each instance $c$ of $C_i$, all values related to $c$ by $R$ are instances of $T_j$ and have a value $v_1$ or...or $v_i$. The domain, $T$, of the value type can be a user defined one, such as \texttt{String}, \texttt{Number}, etc.; recollect that they are values, not objects (hence, not an enumerated class).\\
	{\em example}: {\fontfamily{phv}\selectfont{\small B}}'s values being restricted to one of the strings {\fontfamily{phv}\selectfont{\small \{`a', `b', `c'\}}} and {\fontfamily{phv}\selectfont{\small L}} to integers between {\fontfamily{phv}\selectfont{\small 18}} and {\fontfamily{phv}\selectfont{\small 65}}.
\item \textbf{Mandatory}, $n$-ary predicate with mandatory on role $r_i$ and $i \leq n$:\linebreak $\forall x_i (C_i(x_i) \rightarrow \exists x_1,...,x_{i-1},x_{i+1}, ..., x_n R(x_1, ..., x_n))$\\
	 {\em mapped to}: $C_i \sqsubseteq$ $\exists [r_i]R$\\
	{\em example}: the blob on the {\fontfamily{phv}\selectfont{\small G}} participating in {\fontfamily{phv}\selectfont{\small p6}}.
\item \textbf{Disjunctive mandatory} between the $i$th roles of $n$ different relations, where $n \geq 2$, for $m$-ary relations and $i \leq m$, then $\forall x (C(x) \rightarrow \exists x_1, ..., x_{m-1} \linebreak (R1(x_1,...,x_{i_1-1},x,x_{i_1+1}, ..., x_{m_1}) \lor ... \lor Rn(x_1,...,x_{i_n-1},x, x_{i_n+1}, ..., x_{m_n}) ) ) $\\
	 {\em mapped to}: $C_i \sqsubseteq$ $\sqcup_{i=1}^n \exists [r_j]R_i$ among $n$ relations, each for the $j$th role, $j \leq n$ (for two roles $C_i \sqsubseteq$ $\exists[r_1]R_1 \sqcup$ $\exists[r_1]R_2$)\\
	{\em example}: the first roles of {\fontfamily{phv}\selectfont{\small p16}} and {\fontfamily{phv}\selectfont{\small p17}} to which {\fontfamily{phv}\selectfont{\small N}} participates, i.e., each instance of {\fontfamily{phv}\selectfont{\small N}} must participate either in {\fontfamily{phv}\selectfont{\small p16}} or in {\fontfamily{phv}\selectfont{\small p17}} (or both).
\item \textbf{Uniqueness, $n$:1}, \cite{Halpin89}'s version for binary relation $\forall x,y,z (R(x, y) \land \linebreak R(x, z) \rightarrow y = z)$, which can be generalised to $n$-ary relations\\
	 {\em mapped to}: $C_i \sqsubseteq (\leq 1 [r_i]R)$\\
	 {\em example}: {\fontfamily{phv}\selectfont{\small N}}'s role playing in {\fontfamily{phv}\selectfont{\small p17}}, indicated with the line above the role.\label{it:unique}
\item \textbf{Uniqueness, 1:1}, binary relation, i.e. two times nr.\ref{it:unique}\\ 
	{\em mapped to}: $C_i \sqsubseteq (\leq 1 [r_i]R)$ and $C_j \sqsubseteq (\leq 1 [r_j]R)$\\
	{\em example}: two lines above the two roles in {\fontfamily{phv}\selectfont{\small p22}} and  in {\fontfamily{phv}\selectfont{\small p12}}.\label{nr:1to1} 
\item \textbf{Uniqueness, m:n} on a $n$-ary relation, $n \geq 2$, covering all $n$ roles, is ignored  \cite{Halpin89}: ``repetition of a proposition does not have a logical significance, and is ignored'' \cite{Halpin89}(p4-5) (This is not necessarily true, and will be discussed elsewhere)
, yet the case is included in nr.\ref{it:uniquen} when $i = n$\\
	{\em mapped to}: (\textbf{id} $R [1]r_1, ..., [1]r_i)$, over $i$ roles in $n$-ary relation, $i = n$, and $R$ is a reified relation (see also nr.\ref{it:object})\\
	{\em example}: the line spanning two roles in the binary relation {\fontfamily{phv}\selectfont{\small p2}}.\label{it:uniquemn}
\item \textbf{Uniqueness, $n$-ary} relation where $n \geq 2$, $1 \leq j \leq n$, uniqueness constraint spans at least $n$-1 roles (for it to be elementary)
, and $j$ is excluded from the constraint $\forall x_1,...,x_j,\linebreak...,x_n, y (R(x_1, ...,x_j,..., x_n) \land (R(x_1, ...,y,x_{j+1},...,x_n) \rightarrow x_j = y ) $\\
	 {\em mapped to}: (\textbf{id} $R\mbox{ } [1]r_1, ..., [1]r_i)$ over $i$ roles in $n$-ary relation, $1 \leq i \leq n$, and $R$ is a reified relation (see also nr.\ref{it:object}); note that the FOL formula applies to $n$-1 roles, whereas the $\dlrifd$ one assumes applied to the first $i$ roles ($i \leq n$)\\ 
	 {\em example}: the quaternary relation {\fontfamily{phv}\selectfont{\small p11}} has a uniqueness spanning 3 roles and the ternary {\fontfamily{phv}\selectfont{\small p4}} with a uniqueness over 2 roles.\label{it:uniquen} 
\item \textbf{External uniqueness} (i) among 2 roles $\forall x_1, x_2, y, z (R1(x_1, y) \land R2(x_1, z) \linebreak \land R1(x_2, y) \land R2(x_2, z) \rightarrow x_1 = x_2)$, (ii) among $m$ roles $\forall x_1, x_2, y_1,y_m (\linebreak R1(x_1, y_1) \land ... \land Rm(x_1, y_m) \land R1(x_2, y_1) \land ... \land Rm(x_2, y_m) \rightarrow x_1 = x_2)$\\
	 {\em mapped to}: remodel as $n$-ary relation where $n = m+1$ and a uniqueness over the $n$-1 and use nr.13 (i.e., (\textbf{id} $R\mbox{ } [1]r_1, ..., [1]r_i)$ with $R$ reified), or use the common object type to which the roles relate, i.e., (\textbf{id} $C \mbox{ } [r_1]R_1, ..., [r_1]R_m)$, or, if no such type exist, use a placeholder $C^{\prime}$ instead.\\
	 {\em example}: {\fontfamily{phv}\selectfont{\small F}} is identified by value types {\fontfamily{phv}\selectfont{\small H}} and {\fontfamily{phv}\selectfont{\small I}}, denoted with the encircled line connected to the respective roles in {\fontfamily{phv}\selectfont{\small p7}} and {\fontfamily{phv}\selectfont{\small p8}}.
\item \textbf{Role frequency} with (i) exactly $a$ times, $a \geq 1$, then $\forall x (\exists y_1 R(x, y_1) \rightarrow \exists y_2,...,y_a (y_1 \neq y_2 \land ... \land y_{a-1} \neq y_a \land R(x, y_2) \land ... \land R(x, y_a) )) \land \forall x, y_1, ...,y_{a+1} \linebreak (R(x, y_1) \land ... \land R(x, y_{a+1}) \rightarrow y_1 = y_2 \lor y_1 = y_3 \lor ... \lor y_a = y_{a+1}) $ and (ii) at least $a$ or (iii) at most $a$ times\\
	 {\em mapped to}: (i) $C_i \sqsubseteq (\geq a[r_i]R) \sqcap (\leq a [r_i]R)$ where $a \geq 1$ and (ii) $C_i \sqsubseteq (\geq a[r_i]R)$ and (iii) $C_i \sqsubseteq (\leq a [r_i]R)$\\
	 {\em example}: the $\geq$ {\fontfamily{phv}\selectfont{\small 2}} connected to the role that value type {\fontfamily{phv}\selectfont{\small L}} plays in relation {\fontfamily{phv}\selectfont{\small p10}}.
\item \textbf{Role frequency} with at least $a$ and at most $b$, $1 \leq a$, and $a \leq b$, thus $\forall x (\exists y_1 R(x, y_1) \rightarrow \exists y_2,...,y_a (y_1 \neq y_2 \land ... \land y_{a-1} \neq y_a \land R(x, y_2) \land ... \land R(x, y_a) )) \land \forall x, y_1, ...,y_{b+1} (R(x, y_1) \land ... \land R(x, y_{b+1}) \rightarrow y_1 = y_2 \lor y_1 = y_3 \lor ... \lor y_b = y_{b+1}) $ and for an $n$-ary $R$ where $n>2$ and the amount of $z$ is $n-2$ roles, then 
$\forall x,$ \underline{\emph{z}}$ (\exists y_1 R(x, y_1, $ \underline{\emph{z}}$) \rightarrow \exists y_2,...,y_a (y_1 \neq y_2 \land ... \land y_{a-1} \neq y_a \land R(x, y_2, $\underline{\emph{z}}$) \land ... \land R(x, y_a, $\underline{\emph{z}}$) )) \land \forall x, y_1, ...,y_{b+1} (R(x, y_1, $ \underline{\emph{z}}$) \land ... \land R(x, y_{b+1}, $ \underline{\emph{z}}$) \rightarrow y_1 = y_2 \lor y_1 = y_3 \lor ... \lor y_b = y_{b+1}) $ is \\
	{\em mapped to}: $C_i \sqsubseteq (\geq a[r_i]R) \sqcap (\leq b [r_i]R)$ where $1 \leq a \leq b$ and $i \leq n$\\
	{\em example}: alike the $\geq$ {\fontfamily{phv}\selectfont{\small 2}} of nr.15, but then denoted with, e.g. {\fontfamily{phv}\selectfont{\small 2..5}}.
\item \textbf{Proper subtype}, which holds for subsumption of either object types or of value types, but which cannot be mixed (and note that at times their extensions may contain the same elements) $\forall x (D(x) \rightarrow C(x))$\\
	 {\em mapped to}: $D \sqsubseteq C$; one could add $\neg C \sqsubseteq D$ to ensure that the concepts $D$ and $C$ are never equivalent, but all DL-based reasoners check for subsumption and do classification to detect such distinctions already anyway\\
	 {\em example}: {\fontfamily{phv}\selectfont{\small G}} is a subtype of {\fontfamily{phv}\selectfont{\small F}} and {\fontfamily{phv}\selectfont{\small AB}} of {\fontfamily{phv}\selectfont{\small I}}, denoted with the arrow.
\item \textbf{Subtypes, total (exhaustive) covering} (not formalised in \cite{Halpin89})\\
	 {\em mapped to}: $C \sqsubseteq D_1 \sqcup ... \sqcup D_n$, where the indexed concepts $D$ are subtypes of $C$, in short: $C \sqsubseteq \sqcup_{i=1}^n D_i$\\
	 {\em example}: encircled blob between the two subtype arrows of {\fontfamily{phv}\selectfont{\small V}} and {\fontfamily{phv}\selectfont{\small W}} toward their common supertype {\fontfamily{phv}\selectfont{\small U}}, likewise for value types {\fontfamily{phv}\selectfont{\small AB}}, {\fontfamily{phv}\selectfont{\small AC}} and {\fontfamily{phv}\selectfont{\small I}}.\label{it:tot} 
\item \textbf{Exclusive (disjoint) subtypes} (not formalised in \cite{Halpin89})\\
	 {\em mapped to}: defined among the $1, ..., n$ subtypes of $C$, then $D_i \sqsubseteq \sqcap_{j=i+1}^n \neg D_j$ and $D_i \sqsubseteq C$ for each $i \in \{1, ..., n\}$\\
	 {\em example}: encircled X between the arrows of {\fontfamily{phv}\selectfont{\small T}} and {\fontfamily{phv}\selectfont{\small U}} that are subtypes of {\fontfamily{phv}\selectfont{\small S}}.\label{it:excl}
\item \textbf{Exclusive subtypes, total} (not formalised in \cite{Halpin89})\\
	 {\em mapped to}: use nr.\ref{it:tot} \& nr.\ref{it:excl}\\
	 {\em example}: encircled X with blob on the arrows that subtype {\fontfamily{phv}\selectfont{\small V}} into {\fontfamily{phv}\selectfont{\small X}} and {\fontfamily{phv}\selectfont{\small Y}}.
\item \textbf{Subset over two roles} $r_i$ and $r_j$ in two $n$-ary relations $R_j$ and $R_i$ then for binary $\forall x ( \exists y R_j(x, y) \rightarrow \exists z R_i(x, z)) $ and for an $n$-ary $R$ where $n>2$ and the amount of $w$ is $n-2$ roles, then $\forall x, $ \underline{\emph{w}}$ ( \exists y R_j(x, y, $  \underline{\emph{w}}$) \rightarrow \exists z R_i(x, z, $  \underline{\emph{w}}$)) $\\
	 {\em mapped to}: $[r_i]R_j \sqsubseteq [r_i]R_i$\\
	 {\em example}: the two roles to which {\fontfamily{phv}\selectfont{\small Z}} participate in {\fontfamily{phv}\selectfont{\small p21}} and {\fontfamily{phv}\selectfont{\small p24}}, the latter being a subset of the former.\label{it:ssr} 
\item \textbf{Subset over two $n$-ary relations}, for binary $\forall x, y (R_j(x, y) \rightarrow R_i(x, y))$ and for $n$-ary relation, $\forall x, y (\exists$ \underline{\emph{z}} $(R_j$(\underline{\emph{z}}) $\land x = z_j \land y = z_{j+1}) \rightarrow \exists$ \underline{\emph{w}} $(R_i($\underline{\emph{w}}$) \land x = w_i \land y = w_{i+1}) ) $ 
from \cite{Halpin89}, our compact version $\forall$ \underline{\emph{x}}$(R_1($\underline{\emph{x}}$) \rightarrow R_2($\underline{\emph{x}}$))$\\
	{\em mapped to}: $R_j \sqsubseteq R_i$\\
	{\em example}: {\fontfamily{phv}\selectfont{\small p6}} is a subset of {\fontfamily{phv}\selectfont{\small p5}}; note that the lines connecting the icon to the relation is to the role-divider line instead of in the middle of the role.\label{it:ss}
\item \textbf{Set-equality over two roles} $r_i$ in two binary relations $R_j$, $R_i$ with \linebreak $\forall x ( \exists y R_j(x, y) \equiv \exists z R_i(x, z)) $ and for an $n$-ary $R$ where $n>2$ and the amount of $w$ is $n-2$ roles, then $\forall x, $ \underline{\emph{w}}$ ( \exists y R_j(x, y, $ \underline{\emph{w}}$) \equiv \exists z R_i(x, z, $ \underline{\emph{w}}$)) $\\
	{\em mapped to}: $[r_i]R_j \equiv [r_i]R_i$\\
	{\em example}: an encircled equality sign (as between {\fontfamily{phv}\selectfont{\small p20}} and {\fontfamily{phv}\selectfont{\small p21}}), not drawn. 
\item \textbf{Set-equality over two $n$-ary relations} for binary $\forall x, y (R_j(x, y) \equiv R_i(x, y))$ for $n$-ary relation
$\forall x, y (\exists$ \underline{\emph{z}} $(R_j$(\underline{\emph{z}}) $\land x = z_j \land y = z_{j+1}) \equiv \exists$ \underline{\emph{w}} $(R_i($\underline{\emph{w}}$) \land x = w_i \land y = w_{i+1}) ) $ from \cite{Halpin89}, and our compact version $\forall  \underline{x}(R1(\underline{x})\equiv R2(\underline{x}))$\\
	 {\em mapped to}: $R_j \equiv R_i$\\
	 {\em example}: encircled equality sign between the two relations {\fontfamily{phv}\selectfont{\small p20}} and {\fontfamily{phv}\selectfont{\small p21}}.
\item \textbf{Role exclusion between two roles} $r_i$ and $r_j$ each in $n$-ary relations $R_i$, $R_j$ (which do not necessarily have the same arity), in abbreviated form where $x \in A =_{def} A(x)$, $R_i.r_i$ (resp. $R_j.r_j$) the $r_i$ ($r_j$) role in relation $R_i$ ($R_j$), $1 \leq i \leq n$, then $\forall x \neg(x \in R_i.r_i \land x \in R_j.r_i)$; between $n$ roles $r_1, ..., r_n$ each one in an $m$-ary relation $R_1, ..., R_n$ (which do not necessarily have the same arity) $\forall x \neg((x \in R_1.r_1 \land x \in R_2.r_2) \lor (x \in R_1.r_1 \land x \in R_3.r_3) \lor... \lor (x \in R_{n-1}.r_{n-1} \land x \in R_n.r_n)) $ \\
	 {\em mapped to}: $[r_i]R_i \sqsubseteq \neg [r_j]R_j$ for binary and $([r_1]R_1 \sqsubseteq \neg [r_2]R_2) \sqcup ([r_1]R_1 \sqsubseteq \neg [r_3]R_3) \sqcup ... \sqcup ([r_{n-1}]R_{n-1} \sqsubseteq \neg [r_n]R_n) $ for $n>2$ (and the relations have the same arity)\\
	 	 {\em example}: the encircled cross with lines to each of the three roles to which {\fontfamily{phv}\selectfont{\small M}} participates in relations {\fontfamily{phv}\selectfont{\small p12}}, {\fontfamily{phv}\selectfont{\small p13}} and {\fontfamily{phv}\selectfont{\small p14}}.\label{it:rexnary}
\item \textbf{Relation exclusion between two relations} $R_i$ and $R_j$, $\forall x, y \neg(\exists $\underline{\emph{z}}$ (R_i($\underline{\emph{z}} $ \land x = z_i \land y = z_{i+1}) \land \exists$ \underline{\emph{w}} ($R_j($\underline{\emph{w}}$) \land x = w_j \land y = w_{j+1})  ) $ from \cite{Halpin89}, our version for $n$-ary relations: $\forall \underline{x} (R_1(\underline{x}) \rightarrow \neg R_2(\underline{x})$\\
	 {\em mapped to}: $R_i \sqsubseteq \neg R_j$; note this is relational difference, not negation\\
	 {\em example}: diagrammatic representation 
	  as in previous constraint, but then the connecting lines go to the role-divider lines instead of the middle of the roles.
\item \textbf{Join-subset} among four, not necessarily distinct, relations $R_i$, $R_j$, $R_k$, $R_l$, where $R_i*R_j[c_i, c_j]$ is the projection on columns $c_i$, $c_j$ of the natural join of $R_i$, $R_j$. Then $R_i*R_j[c_i, c_j] \subseteq R_k*R_l[c_k, c_l]$ where the compared pairs must belong to the same type, like \emph{e.g.} $r_i$ of $R_i$ and $r_k$ of $R_k$ is played by $C_a$ and  $r_j$ of $R_j$ and $r_l$ of $R_l$ is played by $C_b$ (See also the example for 3 relations in nr.\ref{it:jeq})\\
	 {\em mapped to}: extend nr.\ref{it:ssr} for subsets of two roles, this $([r_i]R_i \sqcap [r_j]R_j) \sqsubseteq ([r_k]R_k \sqcap [r_l]R_l)$ reduces to query containment  (see \cite{Baader03,Calvanese98})\\
	 {\em example}: the simpler case for three relations is drawn between {\fontfamily{phv}\selectfont{\small p18}} and the relevant roles {\fontfamily{phv}\selectfont{\small R}} and {\fontfamily{phv}\selectfont{\small S}} play in {\fontfamily{phv}\selectfont{\small p17}} and {\fontfamily{phv}\selectfont{\small p16}}, respectively.\label{it:jss}
\item \textbf{Join-equality}, see nr.\ref{it:jss} for notation, then (i) with four distinct relations $R_i*R_j[c_i, c_j] \equiv R_k*R_l[c_k, c_l]$ and (ii) three distinct binary relations $R_i$, $R_j$, $R_k$ such that $\forall x,y (\exists z (R_j(z, x) \land R_k(z, y)) \equiv R_i(x, y) )$\\
	  {\em mapped to}: (i) Extending nr.\ref{it:ssr} for subsets of two roles, this $([r_i]R_i \sqcap [r_j]R_j) \equiv ([r_k]R_k \sqcap [r_l]R_l$) as query containment in both directions, see nr.\ref{it:jss}, and (ii) as simpler version of (i) as $([r_j]R_j \sqcap [r_k]R_k) \equiv ([r_i]R_i \sqcap [r_j]R_i)$\\
	  {\em example}: as in nr.\ref{it:jss}, but then with an encircled equality instead of the encircled subset.\label{it:jeq}
\item \textbf{Join-exclusion}, see nr.\ref{it:jss} for notation, then $R_i*R_j[c_i, c_j] \subseteq \neg R_k*R_l[c_k, c_l]$ (See also the example for 3 relations in nr.\ref{it:jeq}
)\\
	 {\em mapped to}: extending nr.\ref{it:ssr}, then $([r_i]R_i \sqcap [r_j]R_j) \sqsubseteq \neg([r_k]R_k \sqcap [r_l]R_l)$ see also nr.\ref{it:jss}\\
	 {\em example}: not drawn, follows the same pattern as previous two.
\item \textbf{Objectification} (nesting, reification), full uniqueness constraint over the $n$ roles of the $n$-ary relation (note the relaxation described in \cite{Halpin05}), $R_o$ is the objectified relation of $R$, $\forall x (R_o(x) \equiv \exists x_1, ..., x_n (R(x_1, ..., x_n) \land x = (x_1, ...., x_n) ))$\\
	 {\em mapped to}: $R \sqsubseteq \exists [1]r_1 \sqcap (\leq 1 [1] r_1) \sqcap \forall [1] (r_1 \Rightarrow (2: C_1)) \sqcap$ $\exists [1]r_2 \sqcap (\leq 1 [1] r_2) \sqcap \forall [1] (r_2 \Rightarrow (2: C_2)) \sqcap$ ... $\exists [1]r_n \sqcap (\leq 1 [1] r_n) \sqcap \forall [1] (r_n \Rightarrow (2: C_n)) $ where the $\exists [1]r_i$ (with $i \in, \{1, . . . , n\})$ specifies that concept $R$ must have all components $r_1, . . . , r_n$ of the relation $R$, $(\leq 1 [1]r_i )$ (with $i \in \{1, . . . , n\})$ specifies that each such component is single-valued, and $\forall [1](r_i \Rightarrow (2 : C_i ))$ (with $i \in \{1, . . . , n\})$ specifies the class each component has to belong to\\
	 {\em example}: {\fontfamily{phv}\selectfont{\small p9}} is objectified into {\fontfamily{phv}\selectfont{\small FP9J}}.\label{it:object}
\item \textbf{Derived fact type}, implied by the constraints of the roles from which the fact is derived, \emph{i.e.} the original and derived fact type relate through $\leftrightarrow$\\
	 {\em mapped to}: implied by the constraints of the roles from which the fact is derived, hence N/A\\
	 {\em example}: not drawn; the name of the relation is appended with one asterisk instead of the two for derived-and-stored.
\item \textbf{Derived-and-stored} fact type, or conditional derivation, where the predicate indicates that the derivation rule provides only a partial definition of the predicate, \emph{i.e.} the original  and derived fact type relate through $\rightarrow$\\
	 {\em mapped to}: use $\dlrifd$'s \textbf{fd}. With $m$ parameters belonging to the classes $P_1, ... P_m$ (the known part of the partial definition of the predicate) and the result belongs to $R$ (the computed `unknown' part of the partial definition of the predicate), then we have the relation $f_{P_1, ..., P_m}$ with arity $1 + m + 1$, then $f_{P_1, ..., P_m} \sqsubseteq (2 : P_1) \sqcap ... \sqcap (m+1 : P_m) $ with fd as  (\textbf{fd} $f_{P_1, ..., P_m} 1, ..., m+1 \rightarrow m+2$) and the class $C \sqsubseteq \forall[1] (f_{P_1, ..., P_m} \Rightarrow (m + 2 : R))$. Note that for a derivation rule, $m \geq 1$\\
	 {\em example}: {\fontfamily{phv}\selectfont{\small p23$^{**}$}}, i.e., the values for {\fontfamily{phv}\selectfont{\small AA}} are calculated by some formula.

\item \textbf{Role value constraint} (new in ORM2) type $C_i$ only participates in role $r_i$ if an instance has any of the values $\{v_i,...v_k\}$, which is a subset of the set of values $C_i$ can have, for a binary relation, then $\forall x,y ( x \in \{v_i,...,v_k\} \rightarrow  (R(x, y) \rightarrow C_i(x) \land C_j(y)))$ holds\\
	 {\em mapped to}: split the constraint by creating a new subtype $C_i^{\prime}$ for the set of values to which the role is constrained, where the value can be any of $\{v_i,...v_k\}$, and let $C_i^{\prime}$ play role $r_i$, s.t. $C_i^{\prime} \sqsubseteq C_i$ and $C_i^{\prime} \sqsubseteq \forall [r_i]R$ and then use named value types for the value constraints on $C_i^{\prime}$\\
	 {\em example}: only those dates with values {\fontfamily{phv}\selectfont{\small \{01-01-05..01-01-08\}}} can participate when {\fontfamily{phv}\selectfont{\small Z}} participates in {\fontfamily{phv}\selectfont{\small p20}}, although there may be other particular dates recorded for {\fontfamily{phv}\selectfont{\small Z}} that participate in {\fontfamily{phv}\selectfont{\small p19}}; clearly, this role value constraint also holds for {\fontfamily{phv}\selectfont{\small p21}} due to the equality and for {\fontfamily{phv}\selectfont{\small p24}} due to the subsetting.
\end{compactenum}
Arguably, ORM's reference scheme could have been included in the list of constraints. It is depicted in expanded mode for object type {\fontfamily{phv}\selectfont{\small O}}, with a mandatory and 1:1 participation, i.e. that the value type {\fontfamily{phv}\selectfont{\small O\_name}} has unique values by which {\fontfamily{phv}\selectfont{\small O}} is identified, which is also {\fontfamily{phv}\selectfont{\small O}}'s \emph{preferred} reference scheme, indicated with a double line above {\fontfamily{phv}\selectfont{\small O\_name}}'s role (more reference schemes are possible, but one has to choose a preferred one). Although the latter uses a different graphical element, it does not change the logical representation and therefore has not been included in the transformation, above. \\
\indent We demonstrate $\dlrifd$'s representation of ORM's fundamental notion of \emph{fact type} in the following example.
\vspace{-2mm}
\begin{ex}\label{ex:fig3dlr}
The typed ternary relation of Fig. \ref{fig:PatientEx}---an ORM fact type---is represented in {\em $\dlrifd$} as follows: \\
\indent \hspace{2mm} 
{\fontfamily{phv}\selectfont{\small {\em PatientIsAdmittedToHospitalAtDateDate}}} $\sqsubseteq$ \\
\indent \hspace{10mm} 
$(${\fontfamily{phv}\selectfont{\small {\em r1: }}} {\fontfamily{phv}\selectfont{\small {\em Patient}}}$)$  $\sqcap$ $(${\fontfamily{phv}\selectfont{\small {\em haspatients:}}}  {\fontfamily{phv}\selectfont{\small {\em Hospital}}}$)$ $\sqcap$ $(${\fontfamily{phv}\selectfont{\small {\em r3: Date}}}$)$ \\
where the name of the predicate is the one automatically generated by the NORMA software, {\fontfamily{phv}\selectfont{\small {\em PatientAdmittedToHospitalAtDateDate}}}, and the ORM-roles are indexed from left to right, except for the second one, which has as name {\fontfamily{phv}\selectfont{\small {\em haspatients}}}. The mandatory constraint corresponds to \\
\indent \hspace{2mm} {\fontfamily{phv}\selectfont{\small {\em Hospital}}} $\sqsubseteq$ $\exists${\em [}{\fontfamily{phv}\selectfont{\small {\em haspatients}}}{\em ]}{\fontfamily{phv}\selectfont{\small {\em PatientIsAdmittedToHospitalAtDateDate}}}.\\
One can also reify this relation:\\
\indent \hspace{2mm} {\fontfamily{phv}\selectfont{\small {\em PatientIsAdmittedToHospitalAtDateDate}}} $\sqsubseteq$ \\
\indent \hspace{10mm} $\exists [1]r_1 \sqcap (\leq 1 [r1] r_1) \sqcap \forall [r1] (r_1 \Rightarrow (r2:${\fontfamily{phv}\selectfont{\small {\em Patient}}}$)) \sqcap$\\
\indent \hspace{10mm} $\exists [1]r_2 \sqcap (\leq 1 [r1] r_2) \sqcap \forall [r1] (r_2 \Rightarrow (r2:$ {\fontfamily{phv}\selectfont{\small {\em Hospital}}}$)) \sqcap$\\
\indent \hspace{10mm} $\exists [1]r_3 \sqcap (\leq 1 [r1] r_3) \sqcap \forall [r1] (r_3 \Rightarrow (r2:$ {\fontfamily{phv}\selectfont{\small {\em Date}}}$))$\\
\end{ex}
\vspace{-2mm}
The correctness of encoding of this fragment of ORM2, let us call it ORM2$^-$, can be proven by the same line of argumentation as Theorem 6.6 in \cite{Berardi05}: 
\begin{thm}
Let $\mathcal{D}$ be an ORM2$^-$ diagram and $\mathcal{K}_D$ the \emph{$\dlrifd$} knowledge base constructed as described above. Then every instantiation of $\cal{D}$ is a model of $\mathcal{K}_D$, and vice-versa.\label{thm:correnc}
\end{thm}
\begin{proof}
\emph{Both the FOL formalization of the ORM2$^-$ diagram $D$ and the \emph{$\dlrifd$} knowledge base $\mathcal{K}_D$ are over the same alphabet, so their interpretations are compatible. Considering each ORM2$^-$ construct separately as described in items 1-33, above, an interpretation satisfies its FOL formalization if and only if it satisfies the corresponding \emph{$\dlrifd$} assertions. \hfill $\Box$} 
\end{proof}
Like with the results obtained by \cite{Berardi05} for UML Class Diagrams, a consequence of Theorem \ref{thm:correnc} is that reasoning on ORM2$^-$ diagrams can be performed by reasoning on $\dlrifd$ knowledge bases, and, consequently, we obtain Theorems \ref{thm:csat} and \ref{thm:exptime} (analogous to Theorem 6.7 resp 6.8 in \cite{Berardi05}). 
\begin{thm}
Let $\cal{D}$ be an ORM2$^-$ diagram and $\mathcal{K}_D$ the {\em $\dlrifd$} knowledge base constructed as described above. Then an object type $C$ is consistent in $\cal{D}$ if and only if the concept $C$ is satisfiable w.r.t. $\mathcal{K}_D$.\label{thm:csat}
\end{thm}
\begin{thm}
Object type consistency in ORM2$^-$ diagrams is ExpTime-complete.\label{thm:exptime}
\end{thm}
While they are encouraging results for reasoning over ORM2$^-$ diagrams, it is also useful to look at why we have only ORM2$^-$ but not the full ORM2 and what the prospects are for any future extension. This has been discussed extensively in \cite{Keet07dl} and concern the impossibility to represent ORM's ring constraints, i.e., DL role properties, in $\dlrifd$ and certain arbitrary projections that make the language undecidable (such as multi-role frequency, depicted in Fig. \ref{fig:ormconstraints} on {\fontfamily{phv}\selectfont{\small p11}}).

\section{Discussion and related work}
\label{sec:disc}

To assess the merits of the ORM to $\dlrifd$ mapping (or any other DL, for that matter), we first have to address the previous attempt by Jarrar \cite{Jarrar07} and subsequently we will cast a wider scope. 

\subsubsection{Related work.} We deal first with the claimed ORM to $\dlrifd$ ``rules'' with respect to the logic, then with respect to ORM. Jarrar introduces $STRING$ and $NUMBER$ as concepts with the intention to stand in for data types for restricting the values (p190). However, they being primitive concepts, the ``$\{x_1, \ldots, x_n\}$'' denote {\em objects}, not values, i.e., using DL's one-of constructor, which is neither in $\dlrifd$'s syntax nor the intention of {\em value} restrictions, thereby invalidating rules 12 and 13; compare this with nr.2 and nr7, above. For role and relation subset constraints, Jarrar chose to avail of a different language, $DLR$-$Lite$, to deal with projections; it is, however ORM's freedom of allowing arbitrary projects that makes the language undecidable (together with uniqueness constraints, one may regain decidability), hence rule 16 (p191) as such cannot be applied. Further, all mappings and suggestions for ring constraints are incorrect (pp192-193). $\dlrifd$ is a rather poor language for relational properties and one may be better off with $\mathcal{DLR}_{\mu}$ or $\mathcal{SROIQ}$ if ring constraints are one's only interest. More precisely, $\dlrifd$ uses a rewriting for inverses as shown at the end of section \ref{sec:dlrifd}, which is different from the semantic rule $(P^-)^{\mathcal{I}} = \{ (a, b) \in \Delta^{\mathcal{I}} \times \Delta^{\mathcal{I}} \mid (b,a) \in P^{\mathcal{I}} \}$ and therefore the role inclusion with inverse that is needed for symmetry and asymmetry do not have a $\dlrifd$-equivalent (rules 23 and 24). For antisymmetry and irreflexivity,
$\exists R.Self$ is used, which is not in $\dlrifd$ but taken from $\mathcal{SROIQ}$ (note also that antisymmetry in ORM is the normal version, not ``irreflexive antisymmetry'' (i.e., asymmetry) \cite{Horrocks06,Keet07dl}). Also the the role composition operator is not in $\dlrifd$, thereby invalidating rule 26 for the intransitive ring constraint. Last, if one insists on having acyclicity in the language, then $\mathcal{DLR}_{\mu}$ is an option, because there one can represent it using the least/geatest fixpoint operator \cite{Calvanese99mu}. \\
\indent There are three further issues from the ORM perspective as well as DL usage. First, Jarrar introduces a ``$\not\sqsubseteq$'' to capture the notion of {\em proper} subtyping whilst admitting it is not part of the syntax and claiming that ``it can be implemented by reasoning on the ABox to make sure that the population of A and the population of B are not equal'' (p189). However, sub{\em typing} is about the {\em intension} of the concept, not about the {\em extension} (population) at some time (see also \cite{Halpin01} p247). That is, when we have, say, $B \sqsubseteq A$ then a database state where population(B) = population(A) is admissible, be they both empty sets or coincidentally have the same instances; this also means we cannot delegate the reasoning to the ABox as Jarrar proposes. In fact, checking for subtyping is addressed by DL reasoners already. Second, the exclusion constraint in rule 11 (p190) assumes disjointness among two {\em arbitrary} classes, but this is never the case---neither in ORM and ORM2 nor in UML or EER---i.e., disjointness is among {\em subtyped} classes of supertype $C$ (see nr.19). Third, unaries (1-role fact types) cannot be represented in $\dlrifd$ other than by making a binary relation of it with an auxiliary object- or value type. Jarrar tries to solve this by introducing a not further specified concept $BOOLEAN$ where the values ought to be restricted to `true' or `false'; hence, alike mentioned above, making `true' and `false' objects instead of values. If introduced, it is a data type (also called `concrete domain'), but to accommodate ORM's flexibility and suggestion of the alternative notation for unaries (\cite{Halpin01} p83), any new arbitrary $C^{\prime}$ will do, be it an object or value type. For instance, for an unary {\fontfamily{phv}\selectfont{\small Walks}} linked to {\fontfamily{phv}\selectfont{\small  Person}} (i.e., $\forall x (Walks(x) \rightarrow Person(x))$), we could introduce a value type {\fontfamily{phv}\selectfont{\small Walking}} that has domain {\tt String} and values restricted to, say, {\fontfamily{phv}\selectfont{\small `yes'}}, {\fontfamily{phv}\selectfont{\small `no'}}, and {\fontfamily{phv}\selectfont{\small `walkingaid'}}. \\
\indent Concerning elegance in the mappings, there are three points. First, normally the relations contributing to identification in an external uniqueness constraint are directly related to an object type for which the identification is intended, which is also the case in Jarrar's example when one uses the normal compact representation for reference modes. In theory, ORM does not seem to exclude modelling exotic external uniqueness constraints where roles of different relations are combined creatively; however, it is not specifically included in any ORM formalization \cite{Halpin89,Hofstede98} and ORM modelling software does restrict its usage (NORMA prohibits setting the constraint as preferred identifier in a path-based external uniqueness constraint) or does not mention it (FCO-IM). 
To accommodate it nevertheless, one does not need a concept ``$Top$'' to stand in as a natural language version for $\top$ (as in rule 7, p188), because introducing a new placeholder concept $C^{\prime}$ suffices that, if desired, may well be a subtype of another object type (see nr.14). Second, uniqueness on $n$-ary relations is not ideal with {\bf fd} (rule 6) because of the exceptions, and can be done more consistently with {\bf id}; in contrast, {\bf fd}'s are useful in particular for derived and stored relations (UML methods \cite{Berardi05}). Third, the use of $\bot$ in Jarrar's rule 8 for the role frequency constraints of the `at least $a$ and at most $b$' is overly cumbersome, given that a straightforward conjunction ($\sqcap$) suffices (see nr.16, above).

\subsubsection{General discussion.} To put these issues in a broader framework, we observe the following. ORM and ORM2 do have a formal foundation for about 20 years, but when one looks at the details, there is no `standard ORM' like the UML specification \cite{UMLSpec07}. From a logician's perspective, it then seems fair to choose a convenient fragment that fits with one's favourite DL language and define an `ORM$^{\star}$'---or a UML$^{\star}$ or EER$^{\star}$ for that matter---so that each ORM$^{\star}$ diagram (in casu, ORM$^-$) has an equi-satisfiable DL ($\dlrifd$) knowledge base and use the reasoning procedures and complexity results of the chosen language. (This has been done also for UML \cite{Berardi05}, but then with the argumentation that UML being officially informal, one can choose one's own reading of the icons.)  However, it is not the case that any DL language will do just fine. With $\mathcal{DLR}_{\mu}$ or $\mathcal{SROIQ}$ (the basis for OWL 2), we gain regarding the role properties, but loose $n$-ary relations where $n >2$, {\bf id}, and {\bf fd}, hence, also correct objectification, multi-attribute keys, external uniqueness (also called weak entity types or qualified associations), and derived-and-stored fact types (UML methods). Note that the former language is still ExpTime-Complete, but the latter already 2-NExpTime. Clearly, if one takes the  assumption that more features are better, then a formalisation would look different, where one also would be able to add all temporal constraints, deontic constraints, and what have you. This, however, makes practical automated reasoning over conceptual data models unrealistic (more precisely: undecidable). The performance-oriented modeler may want to go to even lower complexity, such as with the $DL-Lite$ family of DLs, and sacrifice even more features compared to the presented mapping so as to stay within NP or NLogSpace complexity (e.g. \cite{Artale07er,Smaragdakis07}). Which features are more important and if one should have many features or less for understandability is a long-standing debate, which we do not want to go into here. There are many options to choose from regarding DLs---the combination of features, complexity, extant implementations---which, perhaps, for a conceptual modeller may be off-putting compared to, say, straightforwardly writing an encoding of a conceptual model as a Constraint Satisfaction Problem (CSP). But note that it is exactly the formal characterisation of a conceptual data modelling language that helps these efforts: if each tool with a constraint-based approach would decide on its own formalizations and CSP encoding, then we would face the situation that each reasoner might come to other derivations, which would not build confidence among the user base of conceptual modellers. Developers of DL-based reasoners, on the other hand, already have a well-established coordination and the reasoners do adhere to the DIG standard [http://dl.kr.org/dig/] that provides uniform access to the DL reasoners such as Fact++, Racer and QuOnto.\\ %
\indent Overall, using a DL, and with respect to conceptual data modelling languages $\dlrifd$ in particular, has several advantages, such as use of a well-studied family of decidable formal languages with model-theoretic semantics, insight in the computational properties, and  availability and active ongoing development of automated reasoners. Another interesting benefit of having carried out the mapping with ORM and $\dlrifd$, is that we now easily can compare it with the mappings from UML and EER to $\dlrifd$ and offer an option to conduct a comparison and unification that could be added to the $\mathcal{DLR}$-based Racer-enhanced proof-of-concept ICOM conceptual modelling tool [www.inf.unibz.it/$\sim$franconi/icom]. It also served to demonstrate that the complex UML qualified associations and association end subsetting---not covered by Berardi et al's mapping \cite{Berardi05}---can easily be represented in $\dlrifd$ as shown in nr.14 and nr.21, respectively. In addition, it can facilitate the investigations into ontological foundations of conceptual data modelling and its languages because there is at least a common, precise, vocabulary and semantics.

\section{Conclusions}
\label{sec:concl}
We have transformed most ORM/ORM2 features into an equivalent $\dlrifd$ representation, and $\dlrifd$ only; thus, any ORM$^-$ diagram has an equi-satisfiable $\dlrifd$ knowledge base. 
This ExpTime-Complete ORM$^-$ suffices for many conceptual data models in practice, i.e, $\dlrifd$ suffices so that there is the benefit of interoperability among EER, UML Class Diagrams and ORM through a common formal language. In addition, by focussing on a very expressive language, we extracted two additional mappings from UML Class Diagrams to $\dlrifd$, being qualified association and subsetting of association ends, and illustrated trade-offs for choosing an adequate DL language.

Current investigation focuses on exploring options to add useful temporal operators and on adding {\bf id} and {\bf fd} to $\mathcal{DLR}_{\mu}$. 


\begin{thebibliography}{10}

\bibitem{Queralt08}
Queralt, A., Teniente, E.:
\newblock Decidable reasoning in {UML} schemas with constraints.
\newblock In: Proc. of CAiSE'08. Volume 5074 of LNCS., Springer (2008)
  281--295

\bibitem{Smaragdakis07}
Smaragdakis, Y., Csallner, C., Subramanian, R.:
\newblock Scalable automatic test data generation from modeling diagrams.
\newblock In: Proc. of ASE'07. (2007)  4--13

\bibitem{Kaneiwa06}
Kaneiwa, K., Satoh, K.:
\newblock Consistency checking algorithms for restricted {U}{M}{L} class
  diagrams.
\newblock In: Proc. of FoIKS '06, Springer Verlag (2006)

\bibitem{Cabot08}
Cabot, J., Claris\'o, R., Riera, D.:
\newblock Verification of {UML}/{OCL} class diagrams using constraint
  programming.
\newblock In: Proc. of MoDeVVA 2008. (2008)

\bibitem{Cadoli07}
Cadoli, M., Calvanese, D., De~Giacomo, G., Mancini, T.:
\newblock Finite model reasoning on {U}{M}{L} class diagrams via constraint
  programming.
\newblock In: Proc. of AI*IA 2007. Volume 4733 of LNAI., Springer (2007)
  36--47

\bibitem{Calvanese98}
Calvanese, D., De~Giacomo, G., Lenzerini, M.:
\newblock On the decidability of query containment under constraints.
\newblock In: Proc. of PODS'98. (1998)  149--158

\bibitem{Artale07er}
Artale, A., Calvanese, D., Kontchakov, R., Ryzhikov, V., Zakharyaschev, M.:
\newblock Reasoning over extended {ER} models.
\newblock In: Proc. of ER'07. Volume 4801 of LNCS., Springer (2007)  277--292

\bibitem{Jarrar07}
Jarrar, M.:
\newblock Towards automated reasoning on {ORM} schemes---mapping {ORM} into the
  {DLR}idf {D}escription {L}ogic.
\newblock In: ER'07. Volume 4801 of LNCS. (2007)  181--197

\bibitem{Franconi00}
Franconi, E., Ng, G.:
\newblock The {I}{C}{O}{M} tool for intelligent conceptual modelling.
\newblock In: Proc. of KRDB'00. (2000) Berlin, Germany, 2000.

\bibitem{Keet07dl}
Keet, C.M.:
\newblock Prospects for and issues with mapping the {O}bject-{R}ole {M}odeling
  language into $\mathcal{DLR}_{ifd}$.
\newblock In: Proc. of DL'07. Volume 250 of CEUR-WS. (2007)  331--338

\bibitem{Berardi05}
Berardi, D., Calvanese, D., De~Giacomo, G.:
\newblock Reasoning on {U}{M}{L} class diagrams.
\newblock Artificial Intelligence \textbf{168}(1-2) (2005)  70--118

\bibitem{UMLSpec07}
{Object Management Group}:
\newblock Superstructure specification.
\newblock Standard 2.1.2, Object Management Group (November 2007)
  http://www.omg.org/spec/UML/2.1.2/.

\bibitem{Halpin01}
Halpin, T.:
\newblock Information Modeling and Relational Databases.
\newblock San Francisco: Morgan Kaufmann Publishers (2001)

\bibitem{Balsters06}
Balsters, H., Carver, A., Halpin, T., Morgan, T.:
\newblock Modeling dynamic rules in {O}{R}{M}.
\newblock In: OTM Workshops 2006--Proc. of ORM'06. Volume 4278 of LNCS.,
  Springer (2006)  1201--1210

\bibitem{Evans05}
Evans, K.:
\newblock Requirements engineering with {O}{R}{M}.
\newblock In: OTM Workshops 2005---Proc. of ORM'05. Volume 3762 of LNCS.,
  Springer (2005)  646--655

\bibitem{Halpin08}
Halpin, T., Morgan, T.:
\newblock Information modeling and relational databases. 2nd edn.
\newblock Morgan Kaufmann (2008)

\bibitem{Pepels05}
Pepels, B., Plasmeijer, R.:
\newblock Generating applications from object role models.
\newblock In: OTM Workshops 2005---Proc. of ORM'05. Volume 3762 of LNCS.,
  Springer (2005)  656--665

\bibitem{Baader03}
Baader, F., Calvanese, D., McGuinness, D.L., Nardi, D., Patel-Schneider, P.F.,
  eds.:
\newblock The Description Logics Handbook.
\newblock Cambridge University Press (2003)

\bibitem{Calvanese01}
Calvanese, D., De~Giacomo, G., Lenzerini, M.:
\newblock Identification constraints and functional dependencies in description
  logics.
\newblock In: Proc. of IJCAI'01. (2001)  155--160

\bibitem{Calvanese98a}
Calvanese, D., De~Giacomo, G., Lenzerini, M., Nardi, D., Rosati, R.:
\newblock Description logic framework for information integration.
\newblock In: Proc. of KR'98. (1998)  2--13

\bibitem{Halpin89}
Halpin, T.:
\newblock A logical analysis of information systems: static aspects of the
  data-oriented perspective.
\newblock PhD thesis, University of Queensland, Australia (1989)

\bibitem{Hofstede93}
Hofstede, A.t., Proper, H., Weide, T.v.d.:
\newblock Formal definition of a conceptual language for the description and
  manipulation of information models.
\newblock Information Systems \textbf{18}(7) (1993)  489--523

\bibitem{Hofstede98}
Hofstede, A.H.M.t., Proper, H.A.:
\newblock How to formalize it? formalization principles for information systems
  development methods.
\newblock Information and Software Technology \textbf{40}(10) (1998)  519--540

\bibitem{Halpin05}
Halpin, T.:
\newblock Objectification of relationships.
\newblock In: Proc. of EMMSAD'05. (2005)

\bibitem{Horrocks06}
Horrocks, I., Kutz, O., Sattler, U.:
\newblock The even more irresistible $\cal{SROIQ}$.
\newblock Proceedings of KR-2006 (2006)  452--457

\bibitem{Calvanese99mu}
Calvanese, D., De~Giacomo, G., Lenzerini, M.:
\newblock Reasoning in expressive description logics with fixpoints based on
  automata on infinite trees.
\newblock In: Proc. of IJCAI'99. (1999)  84--89

\end{thebibliography}

\end{document}